\documentclass[12pt]{iopart}

\expandafter\let\csname equation*\endcsname\relax
\expandafter\let\csname endequation*\endcsname\relax

\usepackage[utf8]{inputenc}
\usepackage[T1]{fontenc}
\usepackage{subfigure}
\usepackage{amsmath,amsfonts,amssymb,amsthm}
\usepackage{algorithm,algpseudocode}
\usepackage{mathtools}
\usepackage{commath} 
\usepackage{comment}
\usepackage{xcolor}
\usepackage{placeins}
\usepackage{booktabs}
\usepackage{hyperref}
\usepackage{braket}

\usepackage{enumitem} 
\newlist{condenum}{enumerate}{1} 
\setlist[condenum]{label=\bfseries Condition \arabic*., 
	ref=\arabic*, wide}

\begin{document}

\title{Molecular Docking via Weighted Subgraph Isomorphism on Quantum Annealers}

\author{Emanuele Triuzzi$^{1,4}$  Riccardo Mengoni$^1$  Francesco Micucci$^4$ Domenico Bonanni$^2$ Daniele Ottaviani$^1$ Andrea Beccari$^3$  Gianluca Palermo$^4$ }
\address{
$^1$CINECA - Quantum Computing Lab - Italy\\
$^2$Università dell'Aquila - Italy\\
$^3$Domp\'e Farmaceutici S.p.A, EXSCALATE - Italy\\
$^4$Politecnico di Milano - Dip. di Elettronica, Informazione e Bioingengeria - Italy
}

\begin{abstract}
Molecular docking is an essential step in the drug discovery process involving the detection of three-dimensional poses of a ligand inside the active site of the protein.  
In this paper, we address the Molecular Docking search phase by formulating the problem in QUBO terms, suitable for an annealing approach.
We propose a problem formulation as a weighted subgraph isomorphism between the ligand graph and 
the grid of the target protein pocket. In particular, we applied a graph representation to the ligand embedding all the geometrical properties of the molecule including its flexibility, and we created a weighted spatial grid to the 3D space region inside the pocket. Results and performance obtained with quantum annealers are compared with classical simulated annealing solvers.
\end{abstract}


\section{Introduction}
In the early stages of drug discovery, computational methods play a critical role in reducing the use of expensive and time-consuming wet lab experiments. By using in-silico techniques, researchers can efficiently filter and prioritize potential drug candidates, significantly reducing the probability of failure in later stages of development. Among these computational approaches, molecular docking \cite{Morris2008, docking06} is an essential step in the drug discovery process as it aims to compute the optimal position and shape of a small molecule, or ligand, when it binds to a target protein \cite{LENGAUER1996402}.
The docking process can be divided into two main tasks. 
The first task is the shape complementarity search \cite{shapematching} that involves the detection of three-dimensional \emph{poses} of the ligand inside the active site of the protein (usually called \emph{pocket}). Meanwhile, the second task is the \emph{binding affinity} evaluation, where the pose is evaluated through a scoring function. 

Due to the computational intensity and resource demands of molecular docking, a variety of techniques have been developed to address these challenges. Furthermore, with the rapid advancements in Quantum Computing \cite{nielsen_chuang_2010} (QC), more of these techniques are increasingly incorporating quantum computing to enhance the accuracy of docking results. 

Quantum computing offers several paradigms for problem-solving, including Quantum Annealing, which leverages quantum tunneling and superposition for combinatorial optimization; Gate-Based Quantum Computing, which uses quantum gates to manipulate qubits; and a Hybrid Classical-Quantum Approach, where a parametric quantum circuit collaborates with a classical optimizer to fine-tune parameters. This paper focuses specifically on shape complementarity search in molecular docking using quantum annealers. In this context, the conventional approach of rigid roto-translation and fragment rotations \cite{GABB1997106} is replaced by framing the problem as a QUBO (Quadratic Unconstrained Binary Optimization) problem.

\begin{figure}[htbp] 
	\centering
	\includegraphics[width=0.9\textwidth]{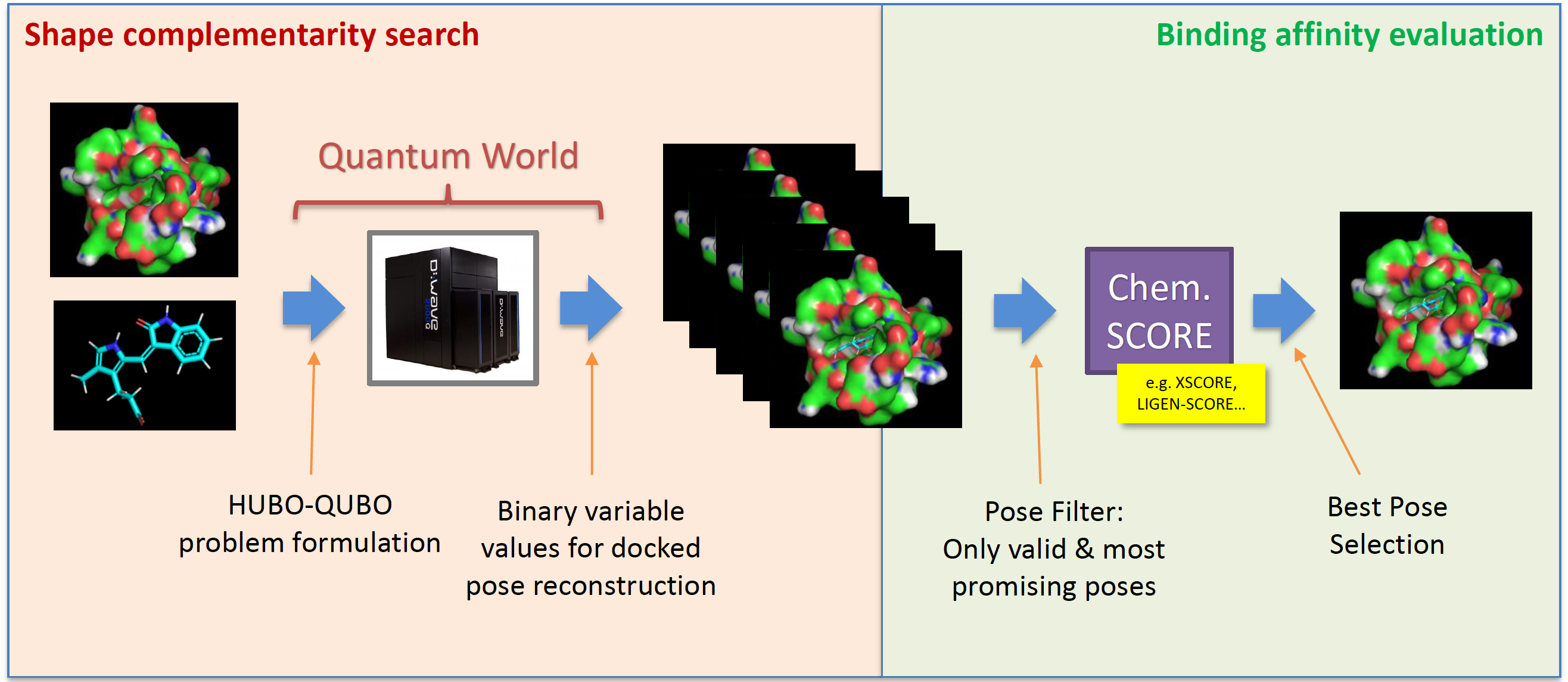} 
	\caption{Complete workflow: the approach addresses the SC search via QA which outputs a sample of valid configurations. The pose filter, in the BA evaluation, selects the most promising poses based on chemical score. }
	\label{fig:Complete workflow.}
\end{figure}
\FloatBarrier

The basic idea behind this approach is to consider interesting docking points within the pocket which identify an active region of the pocket itself.  The number of pocket probes depends on the pocket shape and dimension and are generated using methods derived from the literature (e.g. CAVIAR \cite{CAVIAR}, PASS \cite{Brady2000}, POCASA \cite{POCASA}).  
Probes points can be seen as the vertices of a weighted spatial grid that identifies a certain discretization of the 3D space region inside the pocket, where weights represent distances between probe points.
Ligands, on the other hand, are represented via weighted graphs that embed geometrical properties of the molecule like connectivity between atoms, rotatable bonds, bond length, and values of fixed angles, hence enabling a non-rigid ligand representation.
Finally,  ligand poses are evaluated in terms of an optimal weighted subgraph isomorphism between the ligand graph and the space grid.
This approach has the power of being natively formulated as QUBO problem thus avoiding the wasteful overhead in the number of resources generally associated with the transition from High Order Binary Optimization (HUBO) problem to QUBO (e.g. as required by \cite{Mato_2022}). 

We aimed to develop a purely geometrical Molecular Docking approach and execute it on quantum annealing hardware (D-Wave Advantage and 2000Q), to understand the capabilities of these devices and compare their performances with respect to classical methods like simulated annealing.

\section{Main Related Works}
An increasing number of approaches are now leveraging quantum computers to enhance the drug discovery process \cite{EncMolDoc, doi:10.1126/sciadv.aax1950, ding2024moleculardockingquantumapproximate, mariella2022quantumalgorithmsubgraphisomorphism, Marchand2019, qscreen, Mato_2022, babej2018coarsegrained}. Among these, some of the most prominent focus on addressing the Molecular Docking problem \cite{EncMolDoc, doi:10.1126/sciadv.aax1950, ding2024moleculardockingquantumapproximate, mariella2022quantumalgorithmsubgraphisomorphism}.

One notable method involves reformulating the Molecular Docking problem as a matching problem, where the goal is to map a ligand onto a grid near the target pocket, with the objective function approximating the binding free energy:
\begin{equation}
    \Delta G_{bind} \approx \sum_{i=1}^n w_{i, s_i} \quad s_i = 1, \dots, N
\end{equation}
Here, $n$ represents the number of vertices in the ligand graph, while $N$ indicates the number of vertices in the protein pocket grid. The weight $w_{i, s_i}$ is calculated based on the method used to estimate the binding free energy when ligand vertex $i$ gets mapped on the pocket grid vertex $s_i$. Two examples of these methods are Grid Point Matching (GPM) and Feature Atom Matching (FAM) \cite{EncMolDoc}. This approach is particularly suited for Quantum Annealers, as it can be naturally translated into a Quadratic Unconstrained Binary Optimization (QUBO) problem.

Alternative techniques reformulate the Molecular Docking problem as a maximum weighted clique problem \cite{doi:10.1126/sciadv.aax1950, ding2024moleculardockingquantumapproximate}. In these methods, the ligand and protein are represented as graphs, where the nodes correspond to pharmacophore points - key determinants of a molecule’s pharmacological and biological interactions - and the edges are weighted based on the Euclidean distance between node pairs. These individual graphs are then combined into an interaction graph, where we want to identify the maximum weighted clique. The weights are determined by the pharmacophore point types associated with each vertex. Various techniques can be applied to solve this problem, including Gaussian Boson Sampling \cite{doi:10.1126/sciadv.aax1950} and Variational Quantum Algorithms (VQAs) \cite{ding2024moleculardockingquantumapproximate}.

The Molecular Docking problem can be also solved using a hybrid classical-quantum approach when it is reframed into a Subgraph Isomorphism problem \cite{mariella2022quantumalgorithmsubgraphisomorphism}. In this case, the objective is to identify the presence of a ligand graph within a protein graph. Both graphs are represented by their adjacency matrices, where $A$ corresponds to the protein graph with dimensions $N_A \text{x} N_A$, and $B$ corresponds to the ligand graph with dimensions $N_B \text{x} N_B$. The challenge involves finding a permutation matrix $P$ that minimizes the following cost function:
\begin{equation}
\mathcal{L}_C(P;A,B) = ||SPAP^TS^T - B||^2_F     
\end{equation}
where $S = (\mathcal{I}_{N_B} | \mathcal{O}_{N_B, N_A-N_B})$ and its action on a matrix $A$ (i.e. $SAS^T$) effectively selects the upper $N_B \text{x} N_B$ block of $A$. After embedding the problem into a quantum circuit, they utilize an Ansatz to represent the permutation matrices, parameterized by continuous rotation angles $\theta$. The algorithm alternates between iterations of the stochastic gradient descent (SGD) algorithm, which optimizes the parameters $\theta$ of the Ansatz, and a classical sampler that evaluates several solutions sampled from the superposition of states represented by the current parameters.

\section{Background on Quantum Annealing}
\label{sec:QA}
 Several algorithms can be employed for optimizing a given objective function with several local minima. An example is Simulated Annealing (SA), where thermal energy is supplied to the system in order to escape local minima and a cooling process is used to end up in low-energy states.
Since the transition probability in SA  depends only on the height $ h $ of the barrier separating minima, $ e^{-\frac{h}{k_B T}} $, SA is likely to get trapped in local minima when high barriers are present.

 Similarly, Quantum Annealing (QA)  is a meta-heuristic technique that searches for the global minimum of an objective function by exploiting quantum tunneling and quantum superposition in the exploration of the solutions space.
Since the tunneling probability depends both on the height $ h $ and the width $ w $ of the potential barriers, $ e^{-\frac{w\sqrt{h}}{\Gamma}} $, where $ \Gamma $ is the transverse field strength,  QA may escape local minima in the presence of tall barriers, provided that they are narrow enough.

QA is strictly related to Adiabatic Quantum Computing (AQC), a quantum computation model where a closed quantum system initialized to the ground state of a  simple  Hamiltonian is then adiabatically evolved to reach a desired problem Hamiltonian.  

However,  QA allows fast evolution exceeding the adiabatic regime in an environment with a temperature which is often few mK above absolute zero, hence the annealing process is not guaranteed to converge to the system ground state \cite{7055969}.

\begin{figure}[htbp]
	\centering
	\includegraphics[scale=0.5]{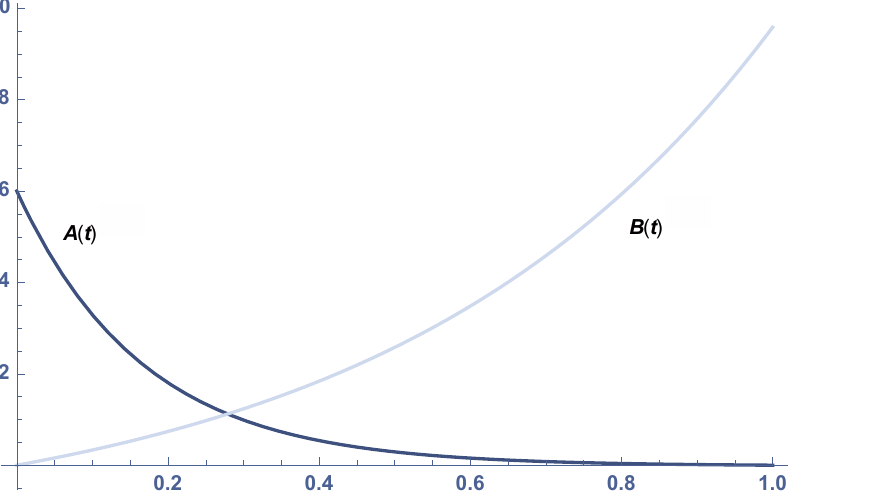}
	\caption{Plot of the  functions  $A(t)$ (blue line) and $B(t)$ (light blue line)   defining the annealing schedule.}
	\label{schedule}
\end{figure}

Mathematically speaking, the  time-dependent Hamiltonian describing   QA is 
\begin{align}
H(t) =  A(t) H_0 + B(t) H_{\rm P}\;,
\label{eq:H}
\end{align}
where $ H_0 $ and   $H_{\rm P}$ are  respectively the initial and problem  Hamiltonian.  The annealing schedule is controlled by  $A(t)$ and $B(t)$, defined in the interval $t \in [0,T_{\rm QA}]$, where $ T_{\rm QA} $ is the total annealing time.
The annealing process is scheduled as follows: at the beginning, the transverse field strength $A(t)$ is large i.e. $A(0) \gg B(0)$  and the evolution is governed by the tunnelling Hamiltonian $ H_0 $;  $A(t)$ and $B(t)$ change in time according to  Fig.\ref{schedule} until  $A(T_{\rm QA}) \ll B(T_{\rm QA})$ where the main term in the evolution is the problem Hamiltonian $ H_P $.
\\

\FloatBarrier

\subsection{D-Wave Quantum Annealer}
\textit{D-Wave Systems} is nowadays the world's leading producer of quantum annealing devices providing access to their machines via the cloud. The current most advanced QA  hardware is a 5000 qubit QPU named \textit{D-Wave Advantage} and a 2048-qubit QPU called \textit{D-Wave 2000Q}.

D-Wave devices are able to find minima of   combinatorial optimization problems  known as  Quadratic Unconstrained Binary Optimization (QUBO) problems of the  form
\begin{equation}
O(x)=\sum_{i}h_i {x}_i+ \sum_{i>j}J_{i,j} x_i x_j
\end{equation}
with $ x_i \in \left\lbrace 0,1 \right\rbrace  $  binary variables and  $ h_i $ and  $ J_{ij} $ parameters with values encoding the optimization problem.

The D-Wave  QA  hamiltonian  $ \mathcal{H}_{QA} $ is given by 
\begin{equation}
\begin{split}
\mathcal{H}_{QA}={A(s)}\sum_{i}\hat{\sigma}^{(i)}_x + {B(s)}H_P
\end{split}
\end{equation}
where annealing parameters $ A(s) $ and $ B(s) $ are  those  shown  in Fig.\ref{schedule}.
The problem Hamiltonian $ H_P $ associated to the objective function $ O(x) $ is expressed   as an  Ising hamiltonian 
\begin{equation}
\begin{split}
H_P=\sum_{i}h_i\hat{\sigma}^{(i)}_z + \sum_{i>j}J_{i,j}\hat{\sigma}^{(i)}_z \hat{\sigma}^{(j)}_z
\label{formula 1}
\end{split}
\end{equation}
where $ \hat{\sigma}^{(i)}_x $ and $ \hat{\sigma}^{(i)}_z $ are Pauli $ x $ and $ z $  operators.

An important aspect to take into account when using the D-Wave annealer is the embedding of the \textit{logical} QUBO problem into the physical architecture of the QPU.
This mapping can be obtained via a heuristic algorithm  (\textit{find\_embedding}), available through the D-Wave Python libraries.
After the embedding, the D-Wave solves the physical problem where logical variables of the QUBO are represented as chains of physical qubits. For this reason, the embedding usually comes with an overhead in the number of resources needed for the optimization.

\section{Molecular Docking Problem Definition}

Molecular docking is a fundamental technique in structure-based drug discovery aimed at accurately positioning a ligand within the binding pocket of a protein. The binding pocket is defined as a cavity (empty space) within the protein structure that provides the spatial constraints necessary for ligand positioning. In the proposed approach, (i) the ligand is represented as a graph capturing either its full molecular structure or a simplified version, (ii) the protein pocket is modelled as a grid, and (iii) the docking process is formulated as a weighted subgraph isomorphism problem mapping the ligand graph onto a subset of the pocket grid.

\label{sec:problem}
\subsection{From Pocket to Space-Grid}
Given a protein pocket, docking points within the pocket are generated using methods derived from the literature \cite{SITEFERRET, CAVIAR, Brady2000, POCASA}.  
Docking points  $v$ are used as vertices of a weighted graph $G_{grid}=\{v,e_{u,v},w_{u,v}\}$ where $e_{u,v}$ are edges connecting points $u$ and $v$ with associated  weights $w_{u,v}$ defined as the Euclidean distance between docking points  $u$ and $v$, i.e.  $w_{u,v}=d(u,v)=|u-v|$. 
The constructed weighted graph $G_{grid}$  identifies a 3D space grid inside the pocket.
It is possible to consider all the edges between pocket points, i.e. $G_{grid}$ is a complete graph, or only a restricted number of edges (e.g. connecting only those edges below a given threshold distance), in this case, $G_{grid}$ is not a complete graph. In the following, we will consider only the first case, where $G_{grid}$ is a complete graph. This is due to the fact that the complexity (i.e. number of linear and quadratic terms) of the QUBO formulation (that will be introduced later on) does not change varying the density of $G_{grid}$. 

\begin{figure}[htbp]
	\centering
\includegraphics[scale=0.35]{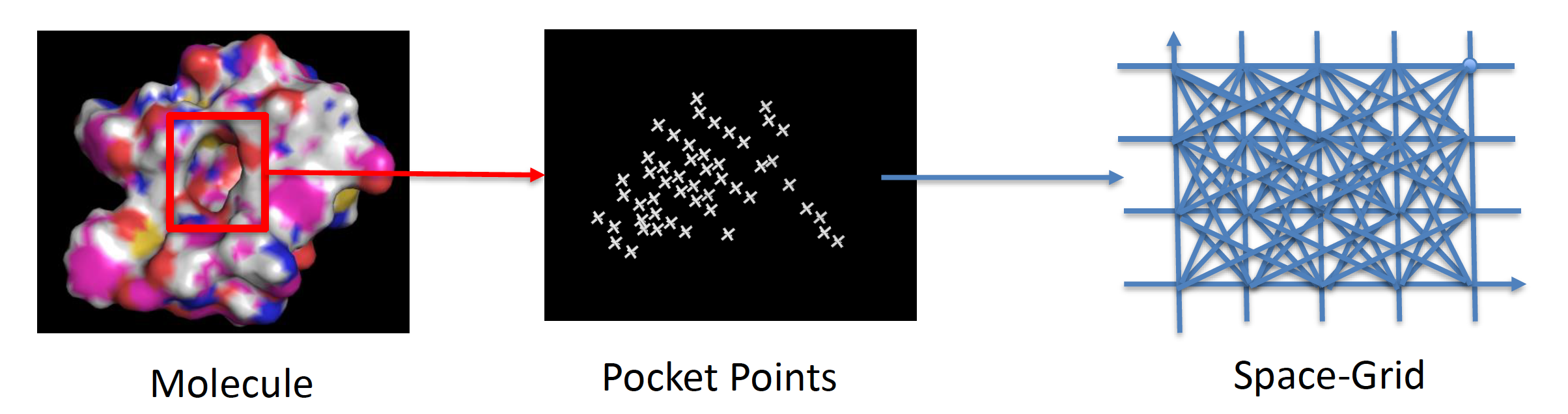}
	\caption{From the molecular pocket we select the pocket points that are used to construct the 3D space grid inside the pocket.}
	\label{grid}
\end{figure}
\FloatBarrier 

\subsection{From Ligand to molecular graph}
In the pre-processing phase, several simplifications are applied to ligands' chemical and geometric properties to align them with the constraints of the current Quantum Computer, specifically in terms of qubits and their interconnections:
\begin{itemize}
    \item Removal of terminal Hydrogen which are external atoms in the ligand with a limited contribution in the overall geometry and shape of the ligand.
    \item Identification of rotatable bonds and fragments, defined as subsets of atoms that are subject to the same rotatable bonds.
    \item Fragment Simplification.  Fragments have been handled in three different ways:\\ a) No fragment simplification, where all atoms belonging to the same fragment are considered.\\ b) Internal fragments (i.e. those fragments connected with more than one rotatable bond) substituted with their center of mass. \\ c) Internal fragments removed, where all atoms belonging to the fragment are removed except those belonging to the shortest path between the rotatable bonds connected to the fragment.
\end{itemize}
While the removal of the hydrogen atoms and the identification of rotatable bonds and fragments are applied to any ligands, fragment simplification defines three different levels of approximation as shown in the scheme below.
\begin{figure}[htbp]
\centering
	\includegraphics[scale=0.12]{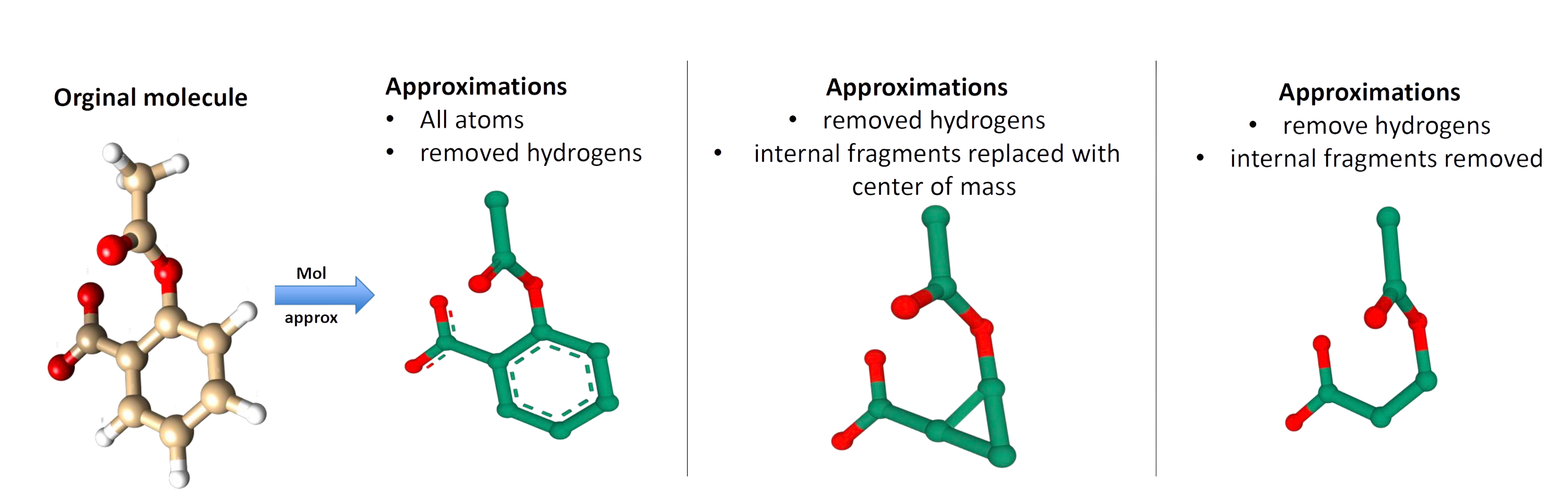}
	\caption{Three different pre-processing approximations: No fragment simplification, Center of mass simplification, and Internal Fragments removal }
	\label{fig:approx}
\end{figure}
\FloatBarrier 

Now that the ligand has undergone such simplification,   information about the geometrical properties of the molecule i.e. connectivity between atoms, rotatable bonds, bonds length, and values of fixed angles are encoded into a weighted graph $G_{mol}=\{i,e_{i,j},w_{i,j}\}$. 
As a starting point, atoms or centers of mass, obtained from the pre-processing step, identify the vertices of the molecular graph $G_{mol}$.
It is necessary now to construct three kinds of edges with related weights  $(e_{i,j},w_{i,j})$   inside $G_{mol}$ having  different roles in the graph representation of the ligand:
\begin{itemize}
    \item Connectivity edges $e^{bond}_{i,j}$ associated with real bonds inside the molecule.  The corresponding weight $w^{bond}_{i,j}$ is exactly the length of the bond (i.e. the  distance between atoms $i$ and $j$).
    \item Fixed bond angles  edges $e^{angle}_{i,j}$. While connectivity edges embed the topological structure of the ligand into the graph, information about fixed angles inside the ligand defining its geometry also needs to be stored in  $G_{mol}$. In Euclidean space, it is possible to fix the angle on-site $\hat{j}$ between two bonds $e^{bond}_{i,j}$ and $e^{bond}_{j,k}$ by adding an ancillary edge $e^{angles}_{i,k}$ between atoms  $i$ and $k$ with proper weight $w^{angle}_{i,k}$ equal to the distance between the two atoms. In this way, we transferred the information about the value of the fixed angle into the length of the edge which is opposite to the angle. 
    \item Fixed dihedral angles edges $e^{dih}_{i,j}$. Based on the construction involving \( e^{bond}_{i,j} \) and \( e^{angle}_{i,j} \), it is possible that the \( G_{mol} \) structure allows rotations around edges that, for chemical reasons, should not be rotatable - because the bond is not of a single type - or because a bond was transformed into a rotatable one due to the ligand approximations introduced. Consequently, it is necessary to add the edges \( e^{dih}_{i,j} \) with appropriate distance weights to fix the dihedral angles for edges that are not rotatable.
\end{itemize}

\begin{figure}[htbp]
\centering
	\includegraphics[scale=0.22]{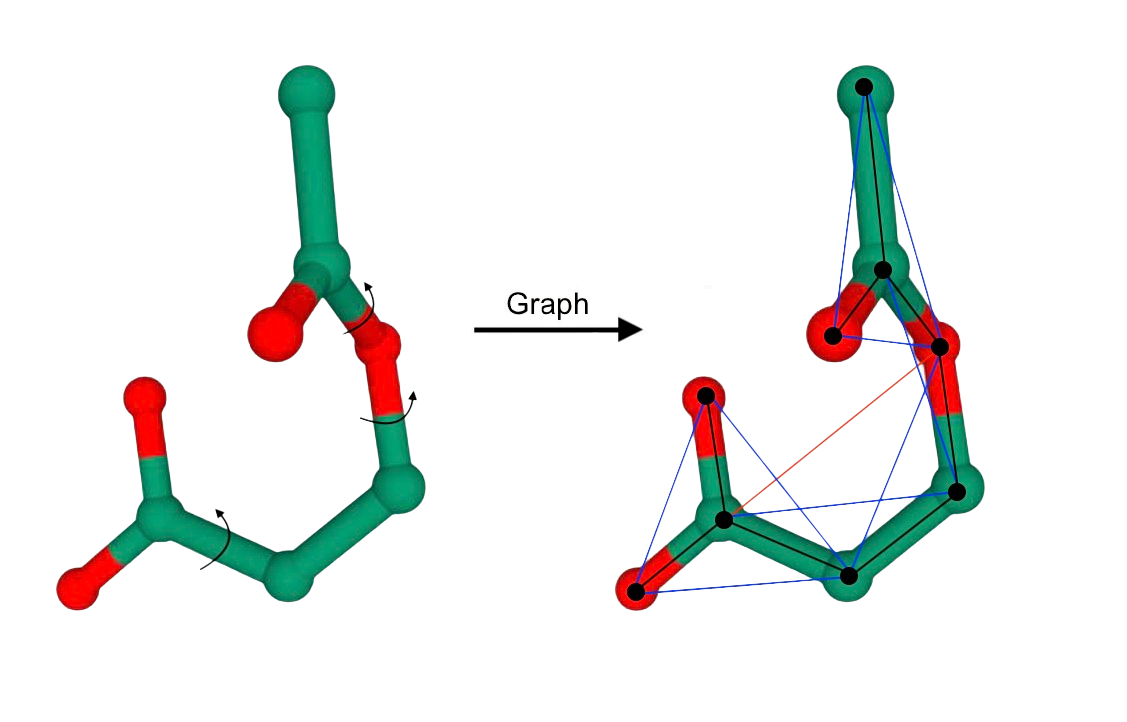}
	\caption{Example of graph construction. On the left, the original molecule obtained after preprocessing, rotatable bonds are identified with arrows. On the right, the graph structure obtained: $e^{bond}_{i,j}$, $e^{angle}_{i,j}$ and $e^{dih}_{i,j}$ are respectively depicted in black, blue and red in the figure.}
	\label{fig:mol_graph}
\end{figure}
\FloatBarrier 

\subsection{Weighted Sub-graph Isomorphism}
The objective of the Molecular Docking problem can be expressed as the optimal matching between the graph associate to the ligand $G_{mol}$ and a sub-graph $G^{mol}_{grid}=\{i',e_{i',j'},w_{i',j'}\}$ of the  space-grid $G_{grid}$  i.e. an injective mapping between vertices $ i\in G_{mol}$ and vertices $ i'\in G^{mol}_{grid}$ such that $G^{mol}_{grid}$ is isomorphic to $G_{mol}$  and      edge weights are minimized:
\begin{equation}
\sum_{i',j'\in G^{mol}_{grid}}(w_{i,j}-w_{i',j'})^2, \ \ \ \ \forall i,j\in G_{mol}
\end{equation}

Hence, ligand poses are evaluated in terms of an optimal weighted subgraph isomorphism between the ligand graph and the space grid. The optimization of the weights enables the search for configurations in which the geometry is maintained, while the subgraph isomorphism allows the ligand to vary its rotatable bonds as well as to roto-translate within the space grid.

This approach has the power of being natively formulated as a QUBO problem thus avoiding the wasteful overhead in the number of resources generally associated with the transition from High Order Binary Optimization (HUBO) problem to QUBO.
Moreover, it does not depend on the initial conditions i.e. initial placement of the ligand and initial values of rotatable bond angles as the graph $ G_{mol}$ encodes only abstract geometrical information about the ligand.

\section{QUBO formulation}

	Given the two graphs $ G_{mol} $ and $ G_{grid}$,  define a binary variable $ x_ {i, i'} $ that  realizes a injective mapping between $ G_{mol} $ and $ G_{grid} $ as follows
	\begin{center}
		$x_{i,i'} = \begin{cases} 1 & \mbox{ if vertex $i \in G_{mol}$ gets mapped to vertex $i' \in G_{grid}$ }  \\ 0 & \mbox{otherwise } \end{cases}$
	\end{center}

	We are now introducing a hard constraint, which must be necessarily satisfied in order to obtain an injective solution. To do so, the following term has to be considered:
	\begin{equation}
	\begin{split}
 H_1=\sum_{i}(1-\sum_{i'}x_{i,i'})^2 
	\label{formula 2}
	\end{split}
	\end{equation}
	At this stage, a second term is added which penalizes a bad mapping i.e. two vertices $i$ and $j$  in $G_{mol}$  connected by an edge, $(i,j)\in G_{mol}$,   mapped into vertices $i'$ and $j'$ in  $G_{grid}$ which instead are not connected, $(i',j')\notin G_{grid}$.
	\begin{equation}
	\begin{split}
     H_2=\sum_{i,j\in{G_{mol}}}\sum_{i',j'\notin{G_{grid}}}x_{i,i'}x_{j,j'}
	\label{formula 3}
	\end{split}
	\end{equation}
Note that $ H_2 $ is not present when the space grid is a fully connected graph.
Using the previously defined terms,  the QUBO formulation of the Sub-graph Isomorphism problem can be expressed as
	\begin{equation}
	\begin{split}
     H_{iso}= H_1+H_2=\sum_{i}(1-\sum_{i'}x_{i,i'})^2  + \sum_{i,j\in{G_{mol}}}\sum_{i',j'\notin{G_{grid}}}x_{i,i'}x_{j,j'}
	\label{formula 3}
	\end{split}
	\end{equation}

Now the optimization term should apply a penalty whenever the geometry is not respected, i.e. edge weights are not preserved in the mapping. For this reason the following QUBO term has to be included.
	\begin{equation}
	\begin{split}
     H_{opt}=\sum_{i,j\in{G_{mol}}}\sum_{i',j'\in{G_{grid}}}(w_{i,j}- w_{i',j'})^2 x_{i,i'}x_{j,j'}
	\label{formula 3}
	\end{split}
	\end{equation}

	The complete QUBO formulation is then
	\begin{equation}
	\begin{split}
	\mathcal{H}_{qubo} =A H_{iso}+  H_{opt} = & A\sum_{i}(1-\sum_{i'}x_{i,i'})^2  + A\sum_{i,j\in{G_{mol}}}\sum_{i',j'\notin{G_{grid}}}x_{i,i'}x_{j,j'} \\ &+B\sum_{i,j\in{G_{mol}}}\sum_{i',j'\in{G_{grid}}}(w_{i,j}- w_{i',j'})^2 x_{i,i'}x_{j,j'}
	\label{formula 4}
	\end{split}
	\end{equation}

	Since $ H_{iso} $ are hard constraints,  parameter $ A $ should be much higher than $ B $ if we want $ H_{iso} $ to be always satisfied (without loss of generality we can set $B=1$ and consider only the $A$ parameter).  However, since in the D-Wave quantum annealer, all the QUBO coefficients (both linear,  $h_i$,  and quadratic, $J_{ij}$) get normalized in the range $[-1,1]$,  it is important to choose $A$ and $B$ in such a way that the normalized coefficients do not become too small.
	For this reason, a good choice  for parameters $A$ and $B$ is the one where
		\begin{equation}
		\min_{i,j,k}\  (h^{iso}_i,J^{iso}_{jk})\geq \max_{i,j,k}\ (h^{opt}_i,J^{opt}_{jk})
	\end{equation}
	where $h^{iso}_i$ and $J^{iso}_{jk}$ are respectively  the linear and quadratic coefficients appearing in $ H_{iso} $ while  $h^{opt}_i$ and $J^{opt}_{jk}$ are   linear and quadratic coefficients of $ H_{opt} $.

\section{Problem Complexity and Embedding}
The total number of binary variables is given by the number of pocket points $N_{grid}$ that define the space-grid $ G_{grid} $  multiplied by the number of nodes $N_{mol}$ in $ G_{mol} $. 
	\begin{equation}
    N_{linear}=N_{mol}\times N_{grid}
	\end{equation}
This number also defines the number of linear terms appearing in the QUBO of the problem. The number of quadratic terms of the QUBO scales  polynomially with  $ N_{linear}$:
	\begin{equation}
    N_{quadratic}\simeq 3N_{mol}\times N_{grid}^2
	\end{equation}

\begin{figure}[htbp]
\centering
	\includegraphics[scale=0.35]{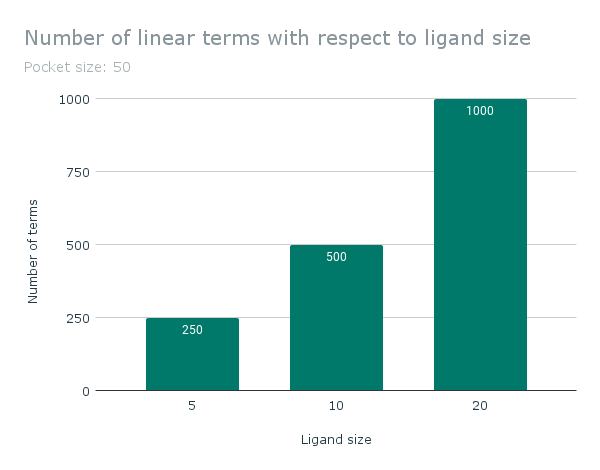}
		\includegraphics[scale=0.35]{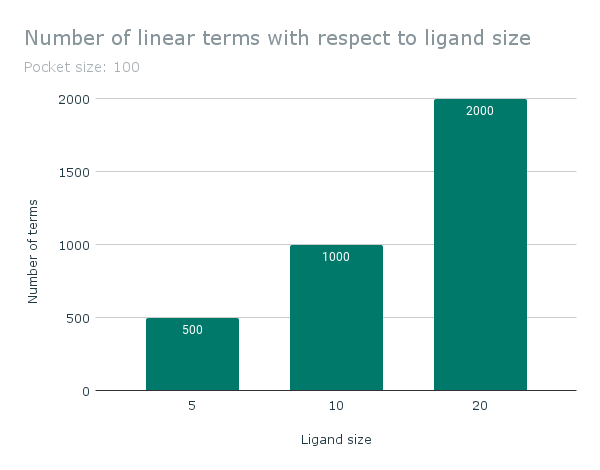}
	\caption{Number of linear terms in the QUBO for a pocket-size with $N_{grid}$ equal to  50 (left) and 100 (right),  increasing ligand size, i.e. number of nodes $N_{mol}$ in $ G_{mol} $  }
	\label{fig:linear_terms}
\end{figure}
\FloatBarrier

\begin{figure}[htbp]
\centering
	\includegraphics[scale=0.35]{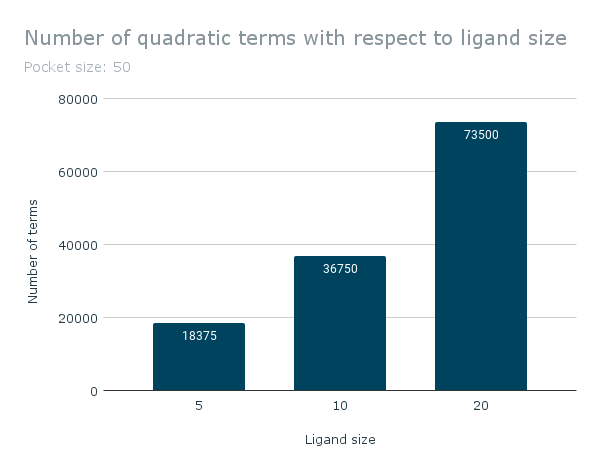}
	\includegraphics[scale=0.35]{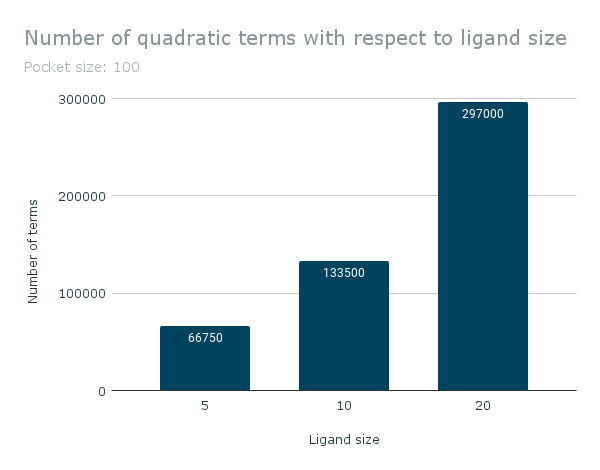}
	\caption{Number of quadratic terms in the QUBO for a pocket-size with $N_{grid}$ equal to  50 (left) and 100 (right),  increasing ligand size, i.e. number of nodes $N_{mol}$ in $ G_{mol} $ }
	\label{fig:quadratic_terms}
\end{figure}
\FloatBarrier 

Given that the formulation of the problem is already in QUBO form, we avoid overhead in the generation of the QUBO, often due to the presence of high-order terms that need to be converted to quadratic and linear terms. Creation time for QUBOs is reported in the plot below and never exceeds 1 minute. To put this number in perspective, the average creation time for the HUBO problem associated to the symbolic approach used in Quantum Molecular Unfolding (Fig.7 of the paper \cite{Mato_2022}) which considered only internal degrees of freedom, was taking up to  1000 seconds for problems with comparable ligand size.

\begin{figure}[htbp]
\centering
	\includegraphics[scale=0.4]{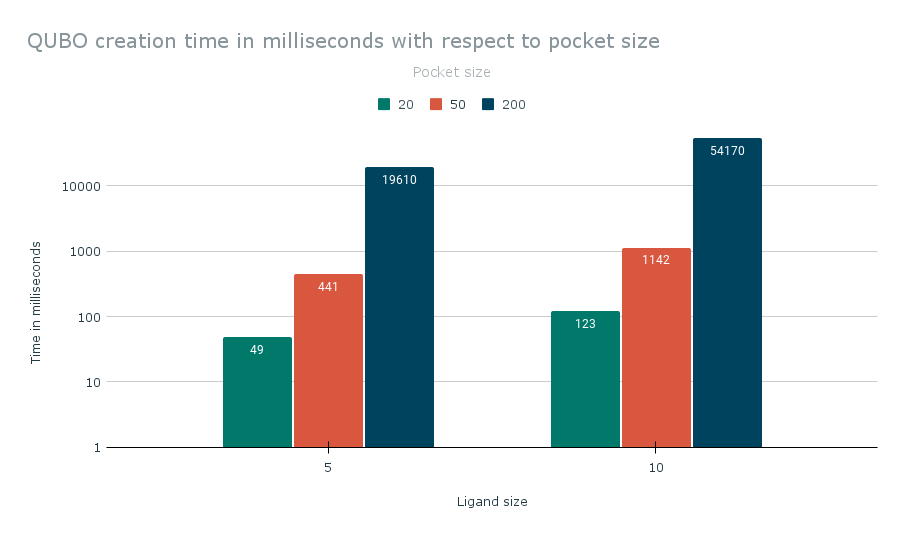}
	\caption{QUBO creation time (in milliseconds) for ligand size of $N_{mol}$ equal to 5 (left) and 10 (right) increasing pocket size $N_{grid}=20, 50, 100$. }
	\label{fig:QUBO_time}
\end{figure}
\FloatBarrier 

\subsection{Embedding into D-Wave hardware}
In the embedding phase, a heuristic algorithm tries to find the optimal matching between problem resources and physical D-Wave hardware.
Several embeddings are possible, so the one that uses the least physical qubits out of a few trials is employed.  It's implicit in this choice that such embedding will also have shortest chains on average, i.e. chains of physical qubits representing a single QUBO variable.

The embeddings shown in the figure below are highly representative of the capabilities of each  QPU topology. 
Regarding the number of qubits, Advantage enables embedding with 2 up to 3 times fewer qubits with respect to 2000Q. 
On average, the chains obtained with 2000Q are 2.6 times longer than those found using the  Advantage topology, which therefore returns shorter chains.

\begin{figure}[htbp]
\centering
	\includegraphics[scale=0.32]{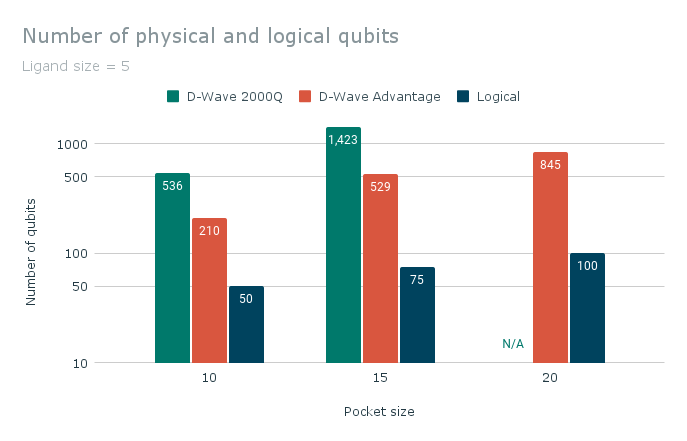}
	\includegraphics[scale=0.32]{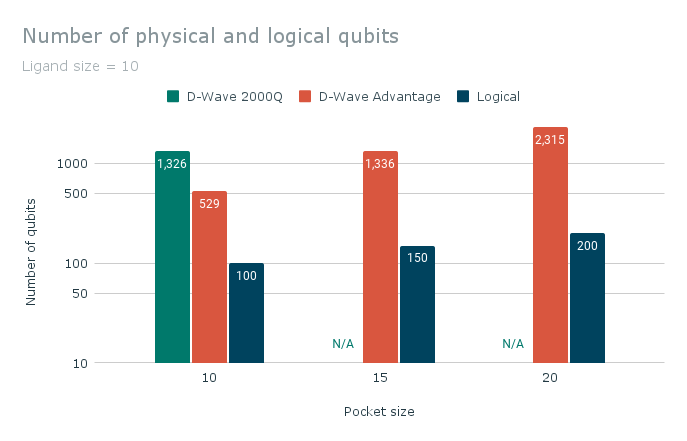}
	\caption{Number of physical qubits obtained after the embedding for ligand size equal to 5 (left) and 10 (right) increasing the size of the pocket space grid. Colors indicate qubits obtained after embedding into D-Wave 2000Q, Advantage compared to the logical qubits before embedding. Embedding in D-Wave 2000Q is limited due to the size of the device which is  able to solve problems up to 100 logical qubits and pocket size 10.}
	\label{fig:qubits}
\end{figure}
\FloatBarrier 
\begin{figure}[htbp]
\centering
	\includegraphics[scale=0.32]{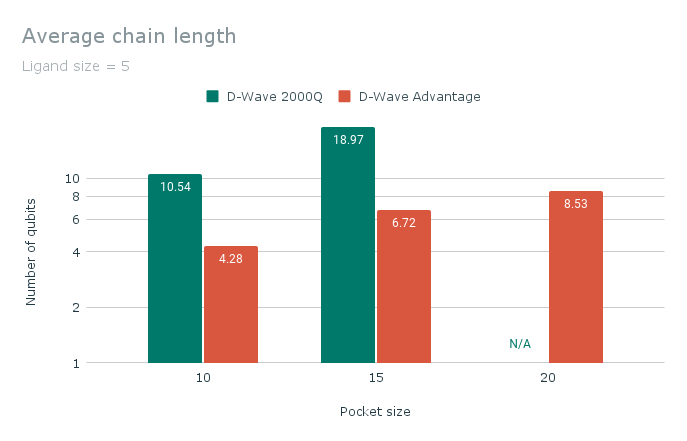}
	\includegraphics[scale=0.32]{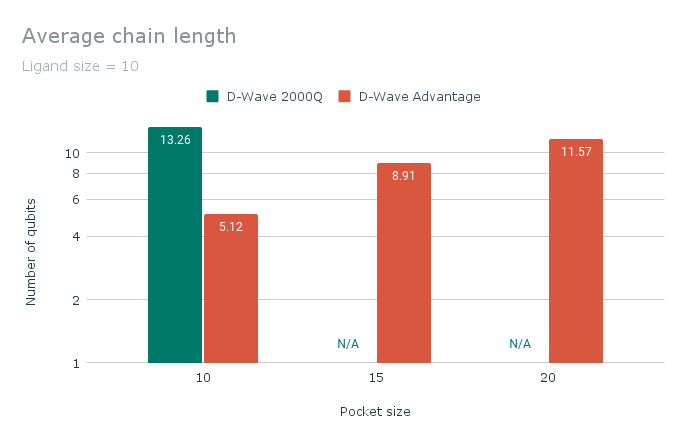}
	\caption{Average number of physical qubits in a chain for ligand size equal to 5 (left) and 10 (right) increasing the size of the pocket space grid. Colors indicate qubits obtained after embedding into D-Wave 2000Q and  Advantage. Embedding in D-Wave 2000Q is limited due to the size of the device.}
	\label{fig:chains}
\end{figure}
\FloatBarrier

\section{Experimental Results}

A short remark should be made about the way annealing algorithms work.
Regarding the QA set-ups, since the QUBO constant \emph{$A$}  modulating the strength of the hard constraint must be tuned, multiple runs for different values assigned to $A$ have been done to identify the most performing one.  $A$ is evaluated as the maximum coefficient in the optimization contribution multiplied by a given term. 
The optimal $A$ parameter has been selected as:
\begin{equation}
    A= 10 \times  \mathrm{max\_opt\_coeff}
\end{equation}

Since we were interested in studying the solution quality by reducing as much as possible the required QPU time, each run is performing $10.000$  cycles of forward anneals,  with an annealing schedule ranging from  $1 \mu sec$ up to $50 \mu sec$.  
Simulated annealing is performed on the same QUBO problems,   with the possibility of choosing the number of epochs and the function that is responsible for the temperature decrease. In our case, the optimal number of annealing epochs was found to be 500 together with a geometrical decrease function of the temperature parameter.
Since the execution of the quantum and simulated annealing are repeated many times in the attempt of finding the best solution, the output would be a sample of the configuration space.

The output sample includes acceptable solutions respecting the hard constraints and minimizing the optimization term of Eq.\ref{opt} as much as possible, that is, solutions that keep the geometry of the ligand as unaltered as possible. 
\begin{equation}\label{opt}
    Opt=\sum_{i',j'\in G^{mol}_{grid}}(w_{i,j}-w_{i',j'})^2, \ \ \ \ \forall i,j\in G_{mol}
\end{equation}

It is important to note that a value equal to zero will not always be obtained in the optimization term since the discretization induced by the space grid does not necessarily allow a perfect matching of the graph associated with the ligand.
For this reason, three thresholds set to $1$, $1.5$, and $2$, related to the value of the optimization term  Eq.\ref{opt}, have been selected.
The best output samples within these thresholds are then saved and used for the calculation of  two metrics: the Average Bond Distortion (ABD) and the Root-Mean-Square Deviation (RMSD) of atomic positions 

The ABD refers to the average square difference between the edge weights of the input molecular graph and those of the output molecule. It is calculated as follows  
\begin{equation}\label{ABD}
\mathrm {ABD} (\mathbf {Ligand_{in}} ,\mathbf {Ligand_{out}} )={\sqrt {{\frac {1}{n_{edges}}}\sum _{(i,j)=1}^{n_{edges}}(w_{i,j}^{in}-w_{i,j}^{out})^2}}
\end{equation}
and expresses the average distortion of each edge of the graph associated with the output ligan with respect to the original one.

The Root-Mean-Square Deviation (RMSD) of atomic positions with respect to the crystal position i.e. the known optimal pose of the ligand inside the pocket. The formula for the RMSD used is the following
\begin{equation}
\mathrm {RMSD} (\mathbf {Ligand} ,\mathbf {Crystal} )={\sqrt {{\frac {1}{n}}\sum _{i=1}^{n}\|l_{i}-c_{i}\|^{2}}}
\end{equation}
where $l_i$ and $c_i$ are corresponding atoms (or fragments) in the ligand and in the crystal respectively. The RMSD metric is essential to measure how far the output sample is from the crystal position.

Moreover, comparing classical algorithms with quantum annealers in terms of absolute time isn't fair, since the quantum devices have run times that are comprised of programming time, readout time, and delay time, besides the time spent in actually performing the annealing.
Therefore, a more suitable metric for comparing annealing solvers is given by the  Time To Solution (TTS),
\begin{equation}\label{eq:TTS}
TTS = \frac{total \ execution \ time}{occurrences \ of  \ best \ solutions}
\end{equation}
The total execution time can be calculated as the number of anneals multiplied by the total access time for a single anneal run. The occurrences of best solutions simply count the number of times a good solution within a given threshold is found in the annealing.
The TTS metric can be interpreted as the inverse of the probability of finding the optimal configuration in a unit time interval. Moreover, the TTS tells us how long we have to wait on average before the annealer outputs a good solution. For this reason, better solvers have low values of TTS.

Results obtained with 2000Q, Advantage, and simulated annealing on simple structures from the PDBbind dataset \cite{PDBbind} are shown in the paragraphs below.

\subsection{Ligand size $N_{mol}=5$, space-grid size $N_{grid}=10$}
\begin{figure}[htbp]
\centering
	\includegraphics[scale=0.45]{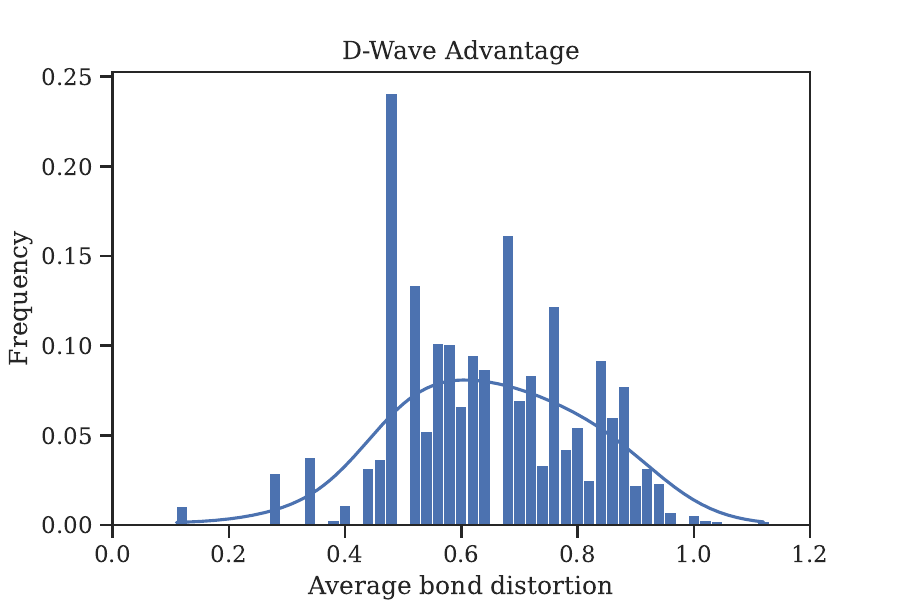}
	\includegraphics[scale=0.45]{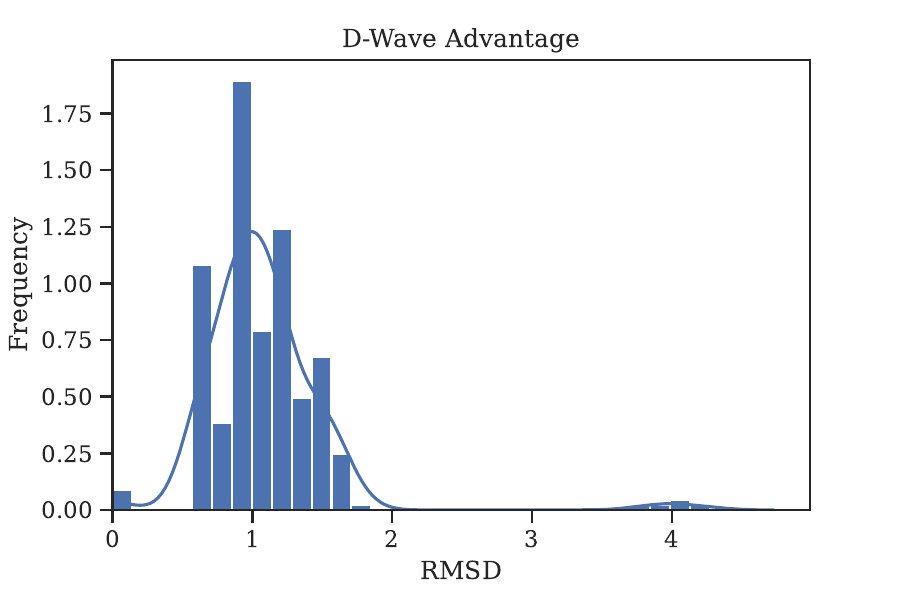}\\
			\includegraphics[scale=0.45]{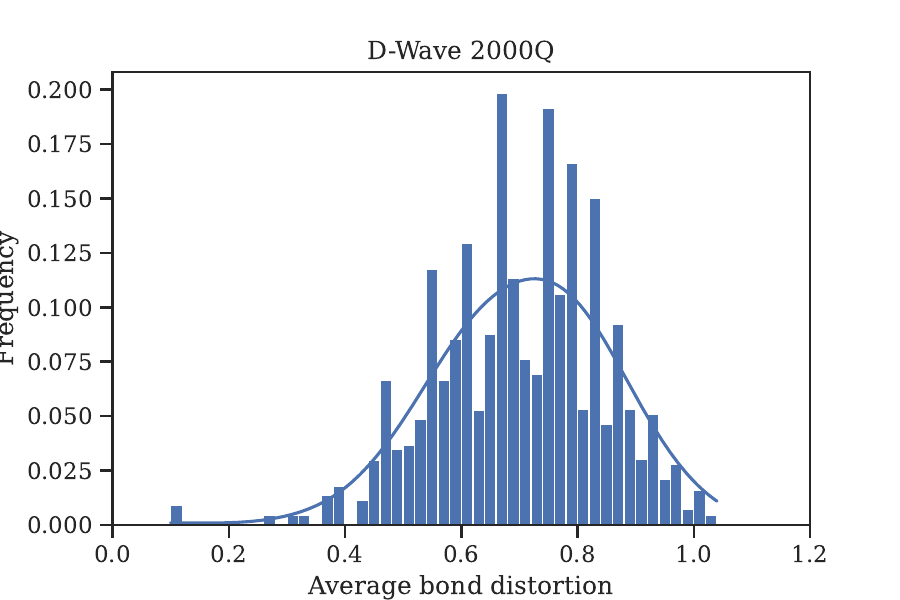}
		\includegraphics[scale=0.45]{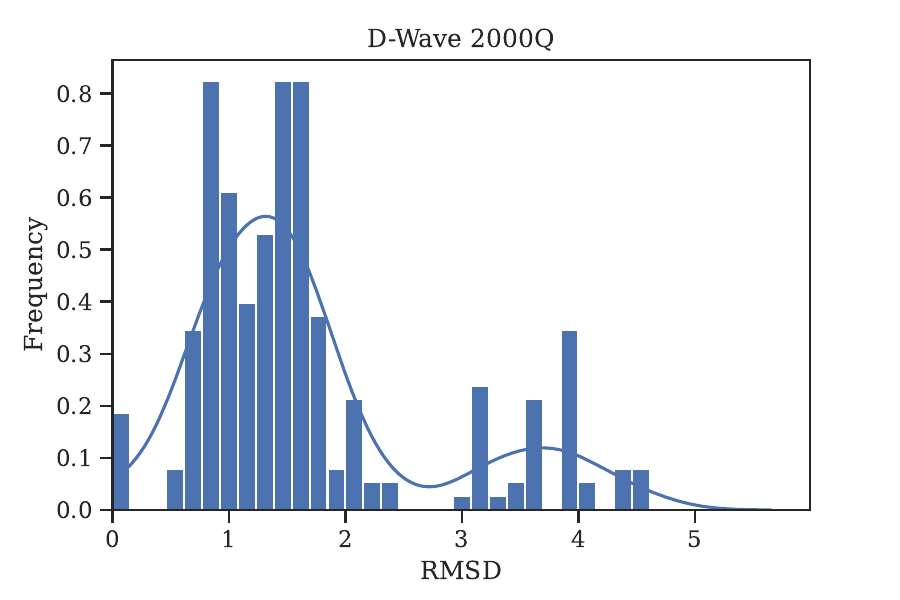}\\
		\includegraphics[scale=0.45]{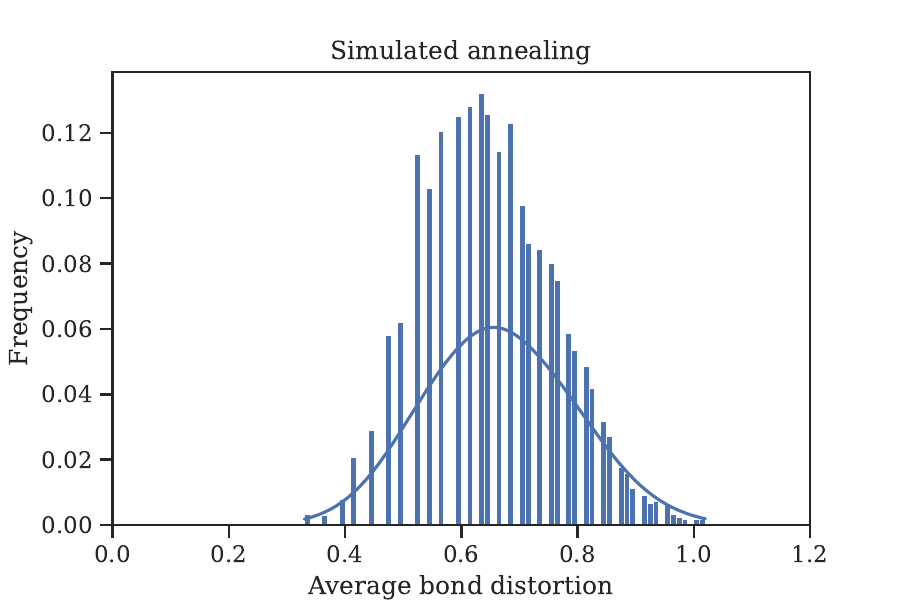}
		\includegraphics[scale=0.45]{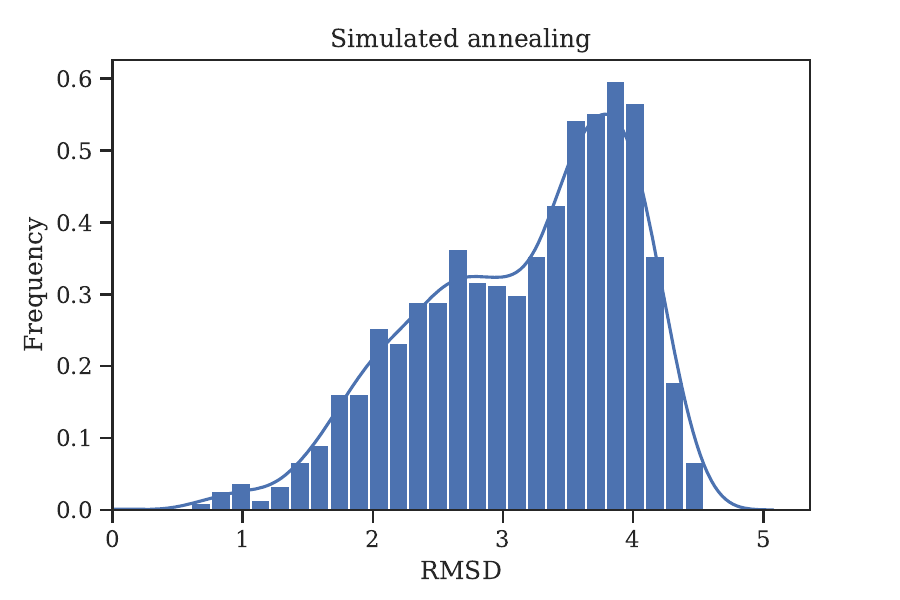}
	\caption{Distribution of solutions in the output sample with respect to ABD (left) and related RMSD (right) obtained from Advantage, DW 2000Q, and Simulated Annealing (SA). The  Total pairwise distance in the plot indicates the value of the optimization term $Opt$. }
	\label{grid_10}
\end{figure}
\FloatBarrier 
The plots in Fig.\ref{grid_10} show the distribution of solutions in the output sample with respect to the ABD and the related RMSD  values obtained from Advantage, DW 2000Q, and Simulated Annealing (SA) solvers. The  ABD  in the plot indicates the value obtained for Eq.\ref{ABD}. By looking at the histograms, it is clear that, despite the largest sample is associated with the SA solver, quantum solvers seem to be able to sample configurations with lower ABD and RMSD with respect to those found by SA. Among the quantum solvers, the DW 2000Q performs better than Advantage wrt RMSD values: this could be explained by the fact that the 2000Q device operates at lower noise levels.
\begin{figure}[htbp]
\centering
	\includegraphics[scale=0.45]{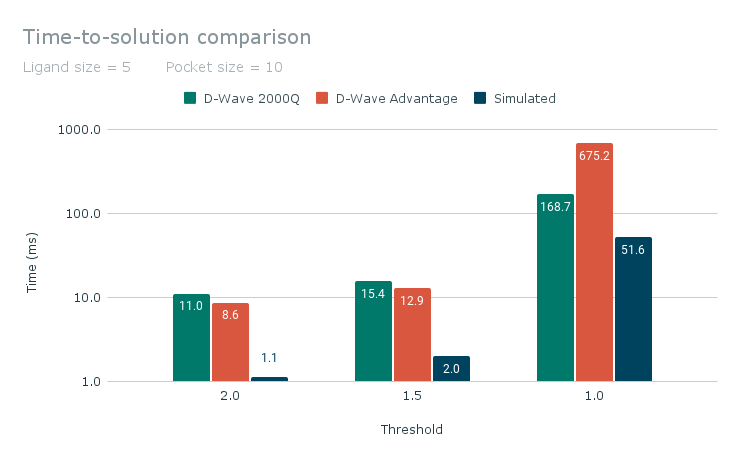}
	\vspace{-.5cm}\caption{Time to solution (in milliseconds) for solutions found below the three threshold values. (Recall that for TTS, the lower the better)}
	\label{fig:tts_5}
\end{figure}
\FloatBarrier 
For what concerns the TTS, solutions below the three thresholds were considered. The SA is able  to find good solutions at very low TTS, if compared to the TTS of the quantum solvers. Anyway, the gap between QPUs and SA is largely reduced when looking for solutions below the lowest threshold. This is also a consequence of the fact that quantum solvers are able to sample fewer valid configurations but with lower scores on average.

\subsection{Ligand size $N_{mol}=5$, space-grid size $N_{grid}=15$}

\begin{figure}[htbp]
\centering
	\includegraphics[scale=0.45]{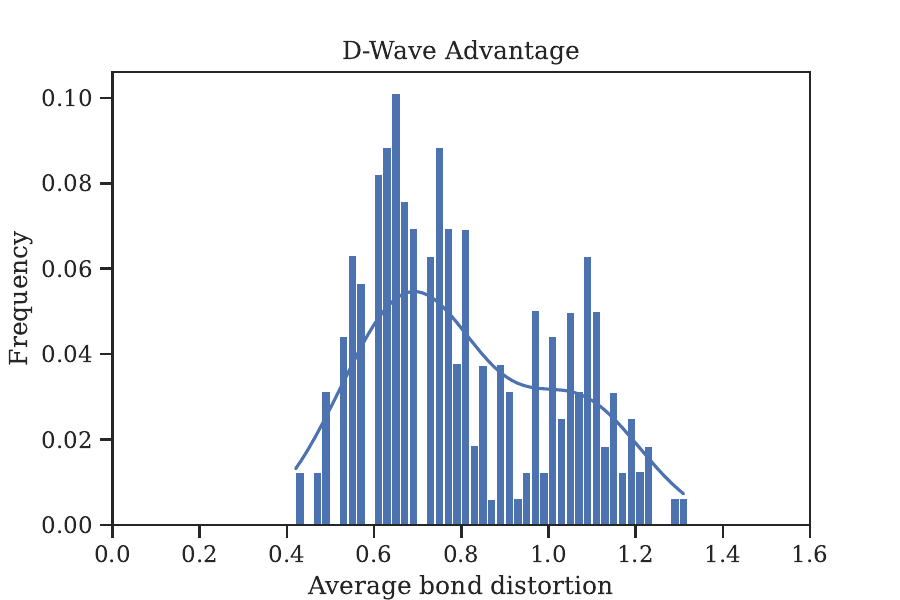}
	\includegraphics[scale=0.45]{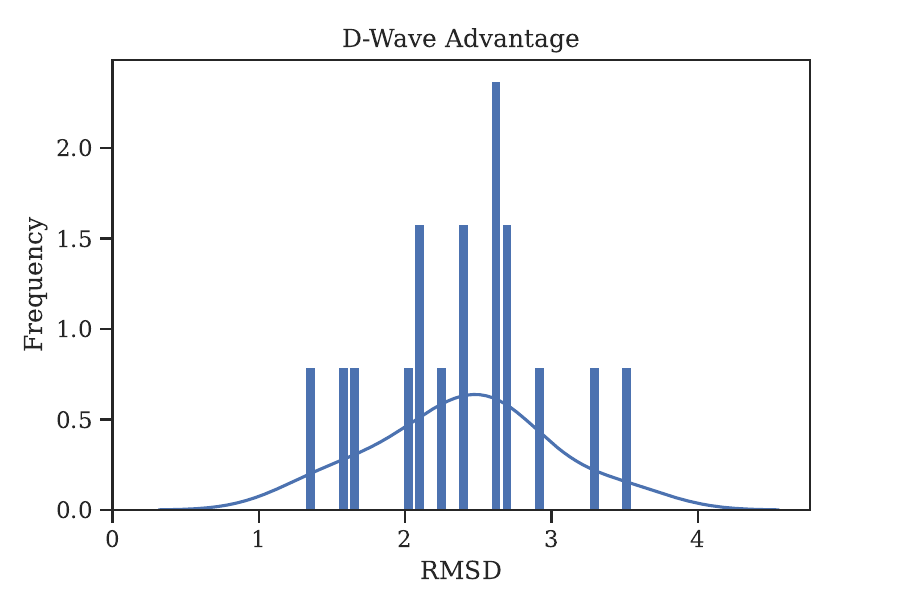}\\
			\includegraphics[scale=0.45]{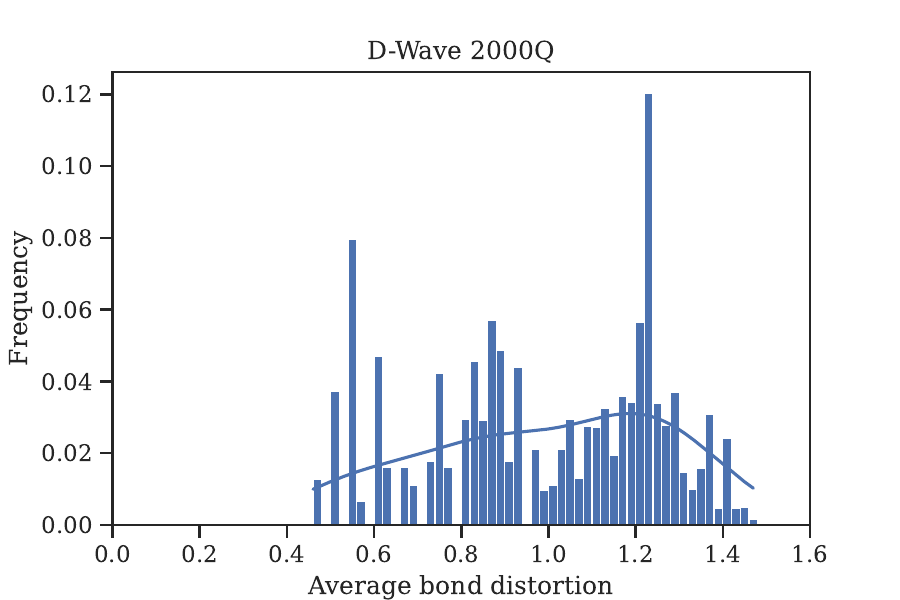}
		\includegraphics[scale=0.45]{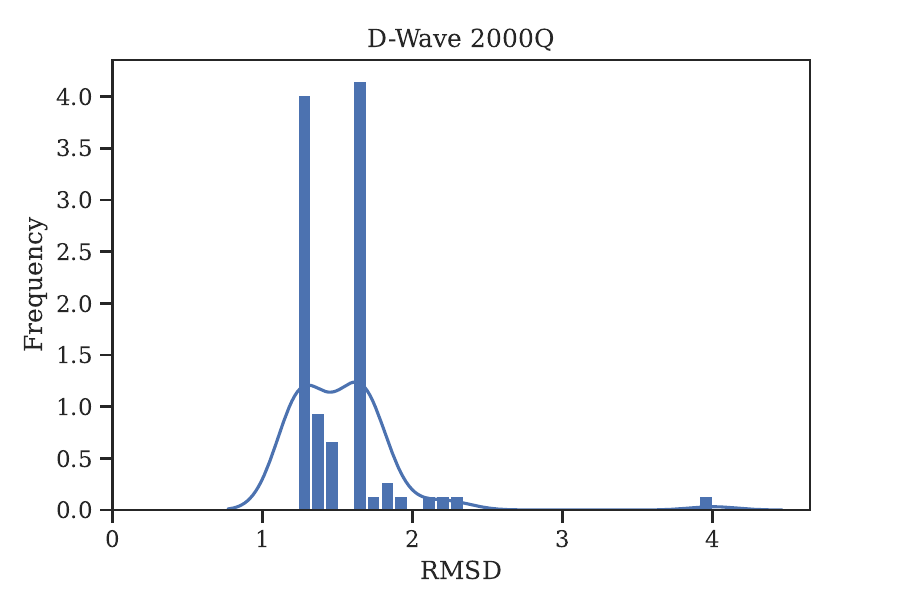}\\
		\includegraphics[scale=0.45]{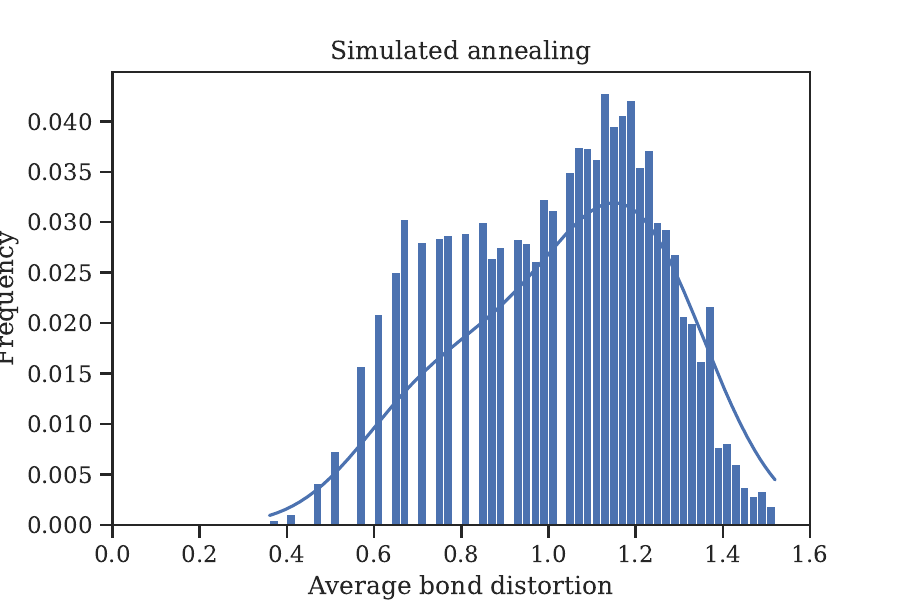}
		\includegraphics[scale=0.45]{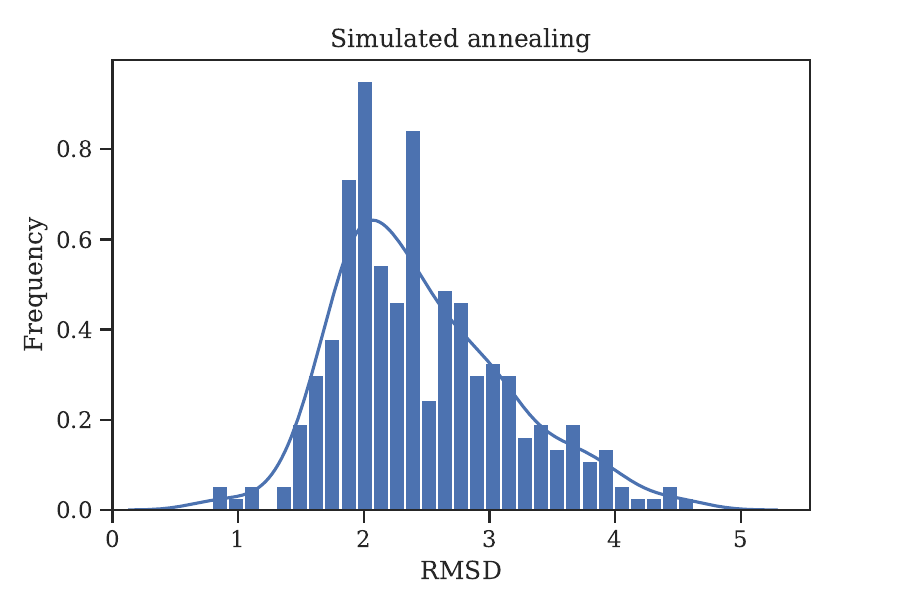}
		\vspace{-.3cm}\caption{Distribution of solutions (left) in the output sample and related RMSD (right) obtained from Advantage, DW 2000Q and Simulated Annealing (SA). }
	\label{grid_15}
\end{figure}

The plots in Fig.\ref{grid_15} show the distribution of solutions in the output sample with respect to ABD and the related RMSD  values obtained from Advantage, DW 2000Q, and Simulated Annealing (SA) solvers. By looking at the histograms, it is clear that the largest sample is associated with the SA solver while quantum solvers (especially DW 2000Q) are  able to sample configurations with low RMSD. However, the number of valid solutions in the samples coming from the QPUs are very limited in size, due to the hardness of satisfying the hard constraint of the QUBO.
\begin{figure}[htbp]
\centering
	\includegraphics[scale=0.45]{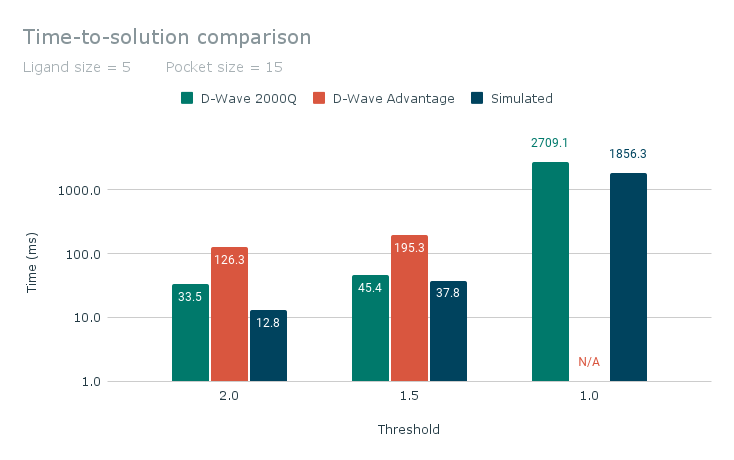}
	\vspace{-.5cm}\caption{Time to solution (in milliseconds) for solutions found below the three threshold values. (Recall that for TTS, the lower the better)}
	\label{fig:tts_5_15}
\end{figure}

For what concerns the TTS, solutions below the three thresholds were considered. The SA is able  to find good solutions at very low TTS, if compared to the TTS of the quantum solvers. The gap between DW 2000Q and SA is small, while Advantage is not able to find solutions below the lowest threshold. This is again due to the fat that 2000Q operates at lower noise levels wrt Advantage.

\section{Conclusions}
In this work, the shape complementarity search phase of the molecular docking phase was addressed by formulating the problem in terms of sub-graph isomorphism so that it can be more easily digested by an annealing approach. 

The basic idea behind our approach is to consider interesting docking points within the pocket which identify an active region of the pocket itself.  
Such points can be seen as the vertices of a weighted spatial grid that identifies a certain discretization of the 3D space region inside the pocket, where weights represent distances between probe points.
Ligands also are represented via weighted graphs that embed geometrical properties of the molecule like connectivity between atoms, rotatable bonds, bond length, and values of fixed angles.
Finally,  ligand poses are evaluated in terms of an optimal weighted subgraph isomorphism between the ligand graph and the space grid. 

This approach has the power of being natively formulated as a QUBO problem thus avoiding the wasteful overhead in the number of resources generally associated with the transition from High Order Binary Optimization (HUBO) problem to QUBO. 
It was shown that QUBO size is polynomial in both the number of nodes of the ligand graph and those of the space grid. Moreover, QUBO creation times are limited to a few seconds or minutes for large problems. 
For what concerns the embedding, we were able to embed problems involving 200 variables and pocket size 20 on the Advantage device while embedding in D-Wave 2000Q is limited due to the size of the QPU  which is able to solve problems up to 100 variables and pocket size 10.
By looking at output samples coming from QPUs and SA solver for a fixed number of anneals (which was set to 10k anneals), it was noticed that  SA can output more valid solutions (i.e. satisfying the hard constraints) w.r.t. the QPUs. However, QPUs seem to be able to sample better quality solutions displaying a lower value of RMSD. This is especially true for D-Wave 2000Q which operates at lower noise levers than Advantage. For what concerns Time to Solution, SA is the best solver, however gap between SA and QPUs decreases when solutions below a small threshold are considered.

\ackn
This work has been supported in terms of resources by CINECA-ISCRA initiative, and by EuroCC-Italy.

\section*{References}
\bibliographystyle{unsrt}

\bibliography{sample.bib}

\begin{thebibliography}{10}

\bibitem{Morris2008}
Garrett~M. Morris and Marguerita Lim-Wilby.
\newblock {\em Molecular Docking}, pages 365--382.
\newblock Humana Press, Totowa, NJ, 2008.

\bibitem{docking06}
Sergio~Filipe Sousa, Pedro~Alexandrino Fernandes, and Maria~Joao Ramos.
\newblock Protein-ligand docking: Current status and future challenges.
\newblock {\em Proteins:Structure, Function, and Bioinformatics}, 65(1):15--26,
  2006.

\bibitem{LENGAUER1996402}
Thomas Lengauer and Matthias Rarey.
\newblock Computational methods for biomolecular docking.
\newblock {\em Current Opinion in Structural Biology}, 6(3):402--406, 1996.

\bibitem{shapematching}
Paul C.~D. Hawkins, A.~Geoffrey Skillman, and Anthony Nicholls.
\newblock Comparison of shape-matching and docking as virtual screening tools.
\newblock {\em Journal of Medicinal Chemistry}, 50(1):74--82, 01 2007.

\bibitem{nielsen_chuang_2010}
Michael~A. Nielsen and Isaac~L. Chuang.
\newblock {\em Quantum Computation and Quantum Information: 10th Anniversary
  Edition}.
\newblock Cambridge University Press, 2010.

\bibitem{GABB1997106}
Henry~A. Gabb, Richard~M. Jackson, and Michael~J.E. Sternberg.
\newblock Modelling protein docking using shape complementarity, electrostatics
  and biochemical information11edited by j. thornton.
\newblock {\em Journal of Molecular Biology}, 272(1):106--120, 1997.

\bibitem{CAVIAR}
Jean-R{\'e}my Marchand, Bernard Pirard, Peter Ertl, and Finton Sirockin.
\newblock Caviar: a method for automatic cavity detection, description and
  decomposition into subcavities.
\newblock {\em Journal of Computer-Aided Molecular Design}, 35(6):737--750,
  2021.

\bibitem{Brady2000}
G.~Patrick Brady and Pieter~F.W. Stouten.
\newblock Fast prediction and visualization of protein binding pockets with
  pass.
\newblock {\em Journal of Computer-Aided Molecular Design}, 14(4):383--401, May
  2000.

\bibitem{POCASA}
Jian Yu, Yong Zhou, Isao Tanaka, and Min Yao.
\newblock {Roll: a new algorithm for the detection of protein pockets and
  cavities with a rolling probe sphere}.
\newblock {\em Bioinformatics}, 26(1):46--52, 10 2009.

\bibitem{Mato_2022}
Kevin Mato, Riccardo Mengoni, Daniele Ottaviani, and Gianluca Palermo.
\newblock Quantum molecular unfolding.
\newblock {\em Quantum Science and Technology}, 7(3):035020, jun 2022.

\bibitem{EncMolDoc}
Jinyin Zha, Jiaqi Su, Tiange Li, Chongyu Cao, Yin Ma, Hai Wei, Zhiguo Huang,
  Ling Qian, Kai Wen, and Jian Zhang.
\newblock Encoding molecular docking for quantum computers.
\newblock {\em Journal of Chemical Theory and Computation}, 19(24):9018--9024,
  2023.
\newblock PMID: 38090816.

\bibitem{doi:10.1126/sciadv.aax1950}
Leonardo Banchi, Mark Fingerhuth, Tomas Babej, Christopher Ing, and Juan~Miguel
  Arrazola.
\newblock Molecular docking with gaussian boson sampling.
\newblock {\em Science Advances}, 6(23):eaax1950, 2020.

\bibitem{ding2024moleculardockingquantumapproximate}
Qi-Ming Ding, Yi-Ming Huang, and Xiao Yuan.
\newblock Molecular docking via quantum approximate optimization algorithm,
  2024.

\bibitem{mariella2022quantumalgorithmsubgraphisomorphism}
Nicola Mariella and Andrea Simonetto.
\newblock A quantum algorithm for the sub-graph isomorphism problem, 2022.

\bibitem{Marchand2019}
D.~J.~J. Marchand, M.~Noori, A.~Roberts, G.~Rosenberg, B.~Woods, U.~Yildiz,
  M.~Coons, D.~Devore, and P.~Margl.
\newblock A variable neighbourhood descent heuristic for conformational search
  using a quantum annealer.
\newblock {\em Scientific Reports}, 9(1):13708, Sep 2019.

\bibitem{qscreen}
Stefano Mensa, Emre Sahin, Francesco Tacchino, Panagiotis~Kl. Barkoutsos, and
  Ivano Tavernelli.
\newblock Quantum machine learning framework for virtual screening in drug
  discovery: a prospective quantum advantage.
\newblock {\em CoRR}, abs/2204.04017, 2022.

\bibitem{babej2018coarsegrained}
Tomas Babej, Christopher Ing, and Mark Fingerhuth.
\newblock Coarse-grained lattice protein folding on a quantum annealer, 2018.

\bibitem{7055969}
Catherine~C. McGeoch.
\newblock {\em Adiabatic Quantum Computation and Quantum Annealing: Theory and
  Practice}.
\newblock 2014.

\bibitem{SITEFERRET}
Luca Gagliardi and Walter Rocchia.
\newblock Siteferret: Beyond simple pocket identification in proteins.
\newblock {\em Journal of Chemical Theory and Computation}, 19(15):5242--5259,
  2023.
\newblock PMID: 37470784.

\bibitem{PDBbind}
Renxiao Wang, Xueliang Fang, Yipin Lu, Chao-Yie Yang, and Shaomeng Wang.
\newblock The pdbbind database: Methodologies and updates.
\newblock {\em Journal of Medicinal Chemistry}, 48(12):4111--4119, 2005.

\end{thebibliography}

\end{document}